\documentclass[letter,traditabstract]{aa} 
\usepackage{graphicx}
\usepackage{natbib}
\usepackage{epsfig}
\usepackage[varg]{txfonts}
\newcommand{\um}{$\mu$m}
\newcommand{\kms}{km~s$^{-1}$}
\newcommand{\cmdue}{cm$^{-2}$}
\newcommand{\cmtre}{cm$^{-3}$}
\newcommand{\lsun}{L$_{\odot}$}
\newcommand{\msun}{M$_{\odot}$}
\newcommand{\jup}{$J_{\rm up}$\,}

\def \arcsec{\hbox{$^{\prime\prime}$}}
\def \arcmin{\hbox{$^{\prime}$}}
\def \degr{\hbox{$^{\circ}$}}

\begin{document}
   \title{The CHESS survey of the L1157-B1 shock: the dissociative jet shock as revealed by Herschel--PACS\thanks{{\it Herschel} is an ESA space observatory with science instruments provided by European-led Principal Investigator consortia and with important participation from NASA.}}

   \author{M. Benedettini\inst{1}, G. Busquet\inst{1}, B. Lefloch\inst{2}, C. Codella\inst{3}, S. Cabrit\inst{4}, C. Ceccarelli\inst{2}, T. Giannini\inst{5}, B. Nisini\inst{5}, M. Vasta\inst{3}, J. Cernicharo\inst{6}, A. Lorenzani\inst{3}, A.M. di Giorgio\inst{1}
          \and the CHESS team         }

   \institute{ 
INAF -- Istituto di Astrofisica e Planetologia Spaziali, via Fosso del Cavaliere 100, 00133 Roma, Italy \email{milena.benedettini@inaf.it}
\and UJF-Grenoble 1/CNRS-INSU, Institut de Plan\'{e}tologie et d'Astrophysique de Grenoble UMR
5274, Grenoble, F-38041, France
\and INAF -- Osservatorio Astrofisico di Arcetri, Largo E. Fermi 5, 50125 Firenze, Italy
\and LERMA, Observatoire de Paris, UMR 8112 du CNRS, ENS, UPMC, UCP, 61 avenue de l'Observatoire, F-75014, Paris
\and INAF -- Osservatorio Astronomico di Roma, Via di Frascati 33, 00040 Monte Porzio Catone, Italy
\and CAB, INTA-CSIC, Department of Astrophysics, Crta Torrej\'{o}n km 4. Torrej\'{o}n de Ardoz, Madrid, Spain
 } 

   \date{Received ; accepted }

 
  \abstract
{Outflows generated by protostars heavily affect the kinematics and chemistry of the hosting molecular cloud through strong shocks that enhance the abundance of some molecules. L1157 is the prototype of chemically active outflows, and a strong shock, called B1, is taking place in its blue lobe between the precessing jet and the hosting cloud. We present the Herschel-PACS 55--210 $\mu$m spectra of the L1157-B1 shock, showing emission lines from CO, H$_2$O, OH, and [\ion{O}{i}]. The spatial resolution of the PACS spectrometer allows us to map the warm gas traced by far-infrared (FIR) lines with unprecedented detail. The rotational diagram of the high-\jup CO lines indicates high-excitation conditions ($T_{\rm ex} \simeq$ 210 $\pm$ 10 K). We used a radiative transfer code to model the hot CO gas emission observed with PACS and in the CO\,(13--12) and (10--9) lines measured by Herschel-HIFI. We derive 200$<T_{\rm kin}<$800 K and $n\geq10^5$ \cmtre. The CO emission comes from a region of about 7\arcsec\, located at the rear of the bow shock where the [\ion{O}{i}] and OH emission also originate. Comparison with shock models shows that the bright [\ion{O}{i}] and OH emissions trace a dissociative J-type shock, which is also supported by a previous detection of [FeII] at the same position. The inferred mass-flux is consistent with the ``reverse'' shock where the jet is impacting on the L1157-B1 bow shock. The same shock may contribute significantly to the high-\jup CO emission.
}

\authorrunning{Benedettini et al.}
   \maketitle
%

\section{Introduction}
Shocks from protostellar outflows heat and compress the surrounding medium, triggering chemical processes that enrich the chemical composition of the environment of young stars \citep[see e.g.][]{bachiller01}. Although outflows and shocks have been studied since the eighties, we are still far from a clear understanding of the interaction between the jet/outflow and the ambient cloud. In particular, studies based on the Infrared Space Observatory (ISO) data have shown that the atomic and molecular lines emitted in the FIR range significantly contribute to the cooling of the shocked material in outflows. However, because of the large ISO beam (80\arcsec) only information averaged along a large part of the outflow lobe could be inferred, and a mixture of both continuous C-type and dissociative J-type shocks has been suggested to be responsible for the excitation of the observed FIR lines \citep[e.g.][]{giannini01,flower09}. 

In the outflow powered by the L1157-mm Class~0 protostar at a distance of 250 pc \citep{looney07}, more than 20 species have been detected in the southern, blue-shifted lobe \citep{bachiller01,arce08}. The brightest shock, B1, has been extensively observed in several molecular species at millimetre \citep{gueth96,gueth98,benedettini07,codella09}, near-infrared \citep{nisini10} and FIR wavelengths \citep{giannini01,nisini10b}, revealing a complex structure with  various shock tracers peaking at different positions. Previous modelling concluded that a (possibly non-stationary) C-type shock is taking place in B1 \citep{gusdorf08, neufeld09,viti11}. 

Because of its chemical richness, L1157-B1 was selected to be observed with the HIFI \citep{degraauw10} and PACS \citep{poglitsch10} spectrometers on board Herschel \citep{pilbratt10}, as part of the Chemical Herschel Survey of Star forming regions (CHESS) Key Program \citep{ceccarelli10}.
The first results obtained by HIFI confirm the chemical richness of B1 \citep{codella10}, including the first detection of HCl in shocked gas \citep{codella11}. They also reveal a high-velocity gas with higher H$_2$O abundance and excitation conditions than the bulk of the emission, and tracing a more compact region within B1 \citep{lefloch10}. 

In this letter we present the PACS spectra of L1157-B1, focusing our analysis on the high-\jup CO, OH and [\ion{O}{i}] lines with the aim of investigating the high-excitation gas component. The H$_2$O lines will be analysed in a forthcoming paper. 
 

\section{Observations and data reduction}

The range spectroscopy mode of the PACS instrument was used to obtain the full range spectrum towards L1157-B1 (RA(J2000) = 20$^{\rm h}$39$^{\rm m}$10\fs2, Dec(J2000) = +68\degr01\arcmin10\farcs5). Two observations were carried out on May 25, 2010, covering the spectral ranges 55--95.2\um\ and 101.2--210 \um. The resolving power ranges between 1000 and 4000, depending on wavelength and order. 
The PACS field of view (FOV) of 47\arcsec $\times$ 47\arcsec\, is composed of 5$\times$5 spatial pixels (spaxels) providing a spatial sampling of 9\farcs4/spaxel, with a position angle of 61\fdg83. Our observations were carried out in staring mode, i.e. our maps are not Nyquist-sampled with respect to the point spread function (PSF).
The data were reduced with the HIPE\footnote{HIPE is a joint development by the Herschel Science Ground Segment Consortium, consisting of ESA, The NASA Herschel Science Center, and the HIFI, PACS and SPIRE consortia.} v5.0 package. The flux calibration was performed by calculating the ratio between the source flux and the telescope background, and subsequently multiplying by the spectrum of Neptune. This procedure gives a flux accuracy of about 10\%  shortwards of 190 \um, while at longer wavelengths the flux calibration is much more uncertain, which caused us to exclude the CO\,(13--12) line at 200.25 \um\, in the analysis. 
Post-pipeline reduction steps were performed to extract the spectrum of each of the 25 spaxels, to calculate the line fluxes with a Gaussian fitting of the line profile and to produce line intensity maps. 
\begin{table}
\caption{List of CO, [\ion{O}{i}] and OH lines detected in spatial pixels centred at offset ($-$5\arcsec,7\arcsec) and (0\arcsec,0\arcsec). For lines covered in the two PACS scans, both measures are reported. Quoted errors ($\pm 1\sigma$) include only statistical noise. Upper limits are 3$\sigma$.}
\label{linelist} 
\centering 
\begin{tabular}{|c|c|c|c|}
\hline
\hline
Line    &$\lambda_{\mathrm{rest}}$ & \multicolumn{2}{c|}{Flux}\\
                       &($\mu$m)   &\multicolumn{2}{c|}{(10$^{-17}$ W m$^{-2}$/spaxel)}\\
        &                          &($-$5\arcsec,7\arcsec) & (0\arcsec,0\arcsec) \\
\hline 
$[\ion{O}{i}]$ $^3P_{1}\--^3P_{2}$&63.18 &50.0$\pm$0.8&13.2$\pm$0.6 \\
OH $^2\Pi_{\frac{1}{2}}-^2\Pi_{\frac{3}{2}}, J={\frac{1}{2}}^--{\frac{3}{2}}^+$ &79.11  &2.6$\pm$0.7& $<$ 1.8 \\
OH $^2\Pi_{\frac{1}{2}}-^2\Pi_{\frac{3}{2}}, J={\frac{1}{2}}^+-{\frac{3}{2}}^-$ &79.18  &2.4$\pm$0.7& $<$  1.8  \\
OH $^2\Pi_{\frac{3}{2}}-^2\Pi_{\frac{3}{2}}, J={\frac{7}{2}}^+-{\frac{5}{2}}^-$ &84.42  &2.0$\pm$0.6& $<$  1.8   \\
OH $^2\Pi_{\frac{3}{2}}-^2\Pi_{\frac{3}{2}}, J={\frac{7}{2}}^--{\frac{3}{2}}^+$ &84.58  &2.4$\pm$0.6& $<$  1.8  \\   
CO (22\--21)                  &118.58 &2.1$\pm$0.2& $<$  1.2   \\
OH $^2\Pi_{\frac{3}{2}}-^2\Pi_{\frac{3}{2}} J={\frac{5}{2}}^--{\frac{3}{2}}^+$ &119.23 &1.9$\pm$0.2 & $<$ 1.2\\   
OH $^2\Pi_{\frac{3}{2}}-^2\Pi_{\frac{3}{2}} J={\frac{5}{2}}^+-{\frac{3}{2}}^-$ &119.44 &2.6$\pm$0.2& $<$ 1.2 \\   
CO (21\--20)                  &124.20  & 2.8$\pm$0.4&1.3$\pm$0.3 \\
CO (20\--19)                  &130.39  & 2.6$\pm$0.2&1.4$\pm$0.2 \\
CO (19\--18)                  &137.23  & 3.3$\pm$0.2&1.8$\pm$0.3 \\
CO (18\--17)                  &144.79  & 3.7$\pm$0.2&1.9$\pm$0.2 \\
$[\ion{O}{i}]$ $^3P_{0}\--^3P_{1}$&145.53 & 2.6$\pm$0.3&0.8$\pm$0.2\\
CO (17\--16)                  &153.28  & 5.2$\pm$0.2&2.2$\pm$0.2 \\
CO (16\--15)                  &162.82  & 6.2$\pm$0.3& 3.0$\pm$0.3 \\
&&6.3$\pm$0.3 & 3.3$\pm$0.2\\
CO (15\--14)                  &173.63  & 7.0$\pm$0.5& 4.1$\pm$0.4 \\
&&6.5$\pm$0.3 & 4.3$\pm$0.3\\
CO (14\--13)                  &186.01  & 7.0$\pm$0.2& 4.7$\pm$0.2 \\
&& 7.2$\pm$0.3& 3.9$\pm$0.2 \\
CO (13\--12)                  &200.25  & 1.8$\pm$0.2$^a$& 1.6$\pm$0.2$^a$ \\
\hline
\end{tabular}
\\
$^a$This line is affected by a large flux calibration error and it is not considered in the analysis.
\end{table}

Table~\ref{linelist} lists the line fluxes of the analysed species, namely CO, OH and [\ion{O}{i}], detected at the centre of the FOV (the nominal position of B1) and towards the PACS emission peak, located at offset ($-$5\arcsec, 7\arcsec) with respect to the centre of the FOV. The spectra of these two spaxels are shown in Figs. \ref{spec2-2} and \ref{spec3-2}.
The spectral range between 161.5 \um\, and 190.2 \um\, was observed in the two observations; the fluxes of the lines detected in both scans generally agree within the errors and in any case never differ by more than 20\%, therefore we consider the averaged value in the following analysis.

\begin{figure}
\centering
\includegraphics[angle=0,width=8cm]{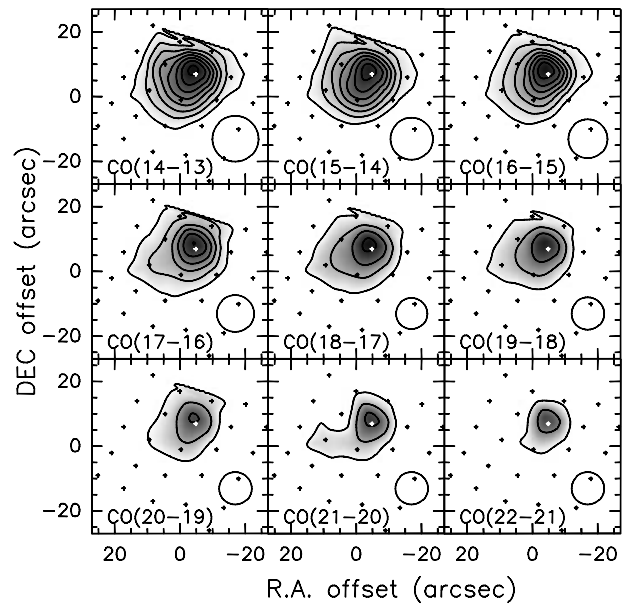}
\caption{Maps of the CO lines from (14--13) to (22--21). First contour and contour steps are 9$\times$10$^{-18}$ W m$^{-2}$/spaxel, corresponding to 3~$\sigma$. Crosses mark the centre of the spaxels of the PACS FOV. Circles represent the half-power beam width.}
\label{mapco}
\end{figure}

\section{Results}

\subsection{Morphology}
\label{morphology}

\begin{figure}
\centering
\includegraphics[angle=0,width=9cm]{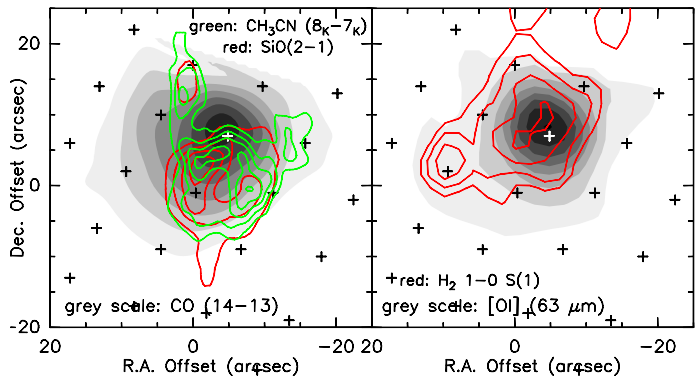}
\caption{\textit{Left:} PACS map of CO\,(14--13) in L1157-B1 (gray scale) compared with SiO\,(2--1) (red contours) and CH$_3$CN (8$_K$--7$_K$) (green contours). \textit{Right:} PACS map of [\ion{O}{i}]63\um\,(gray scale, first level is 3$\times$10$^{-17}$ W m$^{-2}$/spaxel, corresponding to 3~$\sigma$, steps are 7$\times$10$^{-17}$ W m$^{-2}$/spaxel) compared with H$_2$\,$v=$1-0\,S(1) (red contours). Crosses mark the centre of the spaxels of the PACS FOV.}
\label{comparison}
\end{figure}

Figure~\ref{mapco} presents the maps of the CO lines from (14--13) to (22--21).  The emission peaks towards spaxel ($-$5\arcsec,7\arcsec), with an elongation towards south-east.
The deconvolved full with at half maximum (FWHM) size of the CO peak, estimated by subtracting in quadrature the  FWHM of the PSF, ranges between 5\arcsec and 10\arcsec\, depending on wavelength, with a mean value of 7\arcsec, hence marginally resolved in these observations. The emission from [\ion{O}{i}] and OH peaks in the same spaxel as high-\jup CO.
In Fig.~\ref{comparison} we show the CO\,(14--13) and [\ion{O}{i}] 63 \um\, PACS maps superimposed on tracers of postshock gas such as SiO\,(2--1) \citep{gueth98} and CH$_3$CN (8$_K$--7$_K$) \citep{codella09} observed with the Plateau de Bure at a resolution of $\sim$ 3\arcsec.  While both SiO and CH$_3$CN trace an extended bow shock at the tip of the outflow cavity traced by CO\,(1--0) \citep{gueth96}, it is clear that high-\jup\ CO, OH and [\ion{O}{i}] are peaking in a different region, at the rear of this bow shock.
On the other hand, [\ion{O}{i}], OH and high-\jup CO lines correlate well with the [\ion{Fe}{ii}] at 26 \um\, \citep{neufeld09}, the H$_2$ ro-vibrational line $v=$1-0\,S(1) \citep{caratti06} that traces hot gas ($T>$2500~K) as well as with the pure H$_2$ rotational lines tracing gas at temperatures between 200 and 1400 K \citep{nisini10}. This indicates that the FIR PACS lines trace a more compact and more excited region than the bulk of the B1 bow shock, most probably the ``jet-shock'' where the jet powered by the protostar is impacting on the internal cavity wall.

\subsection{Physical conditions in the high-\jup\ CO emitting gas}

\begin{figure}
\centering
\includegraphics[angle=90,scale=0.36]{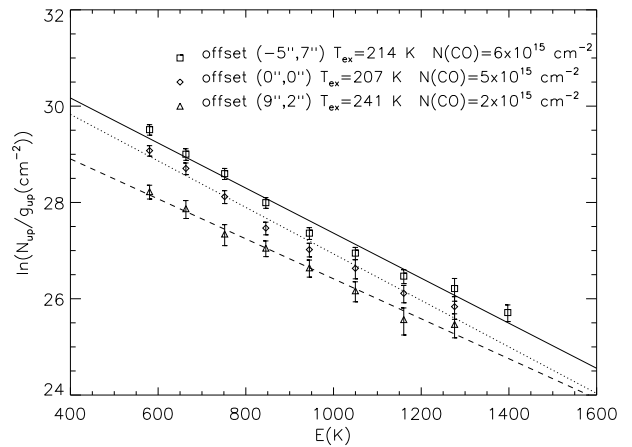}
\caption{CO rotational diagrams at 13" resolution, with superimposed linear fits. Diamond, squares and triangles correspond to offset (0\arcsec,0\arcsec), ($-$5\arcsec,7\arcsec) and (9\arcsec,2\arcsec), respectively. The inferred excitation temperatures and total CO column densities are indicated.}
\label{co_rotdiagram}
\end{figure}

In order to look for possible spatial gradients of the physical conditions in the CO gas, we calculated the excitation temperatures, which gives a lower limit to the kinetic temperature if the gas is not in local thermodynamic equilibrium (LTE). To correct for the different beam sizes, we convolved all CO line maps to the resolution of the transition with the longest wavelength, i.e. 13\farcs1 (PSF at 186 \um) and then extracted the fluxes at the position of each spaxel. The fluxes for spaxels ($-$5\arcsec,7\arcsec) and
(0\arcsec,0\arcsec) are reported in Table \ref{flux_lvg}. We then constructed the corresponding rotation diagrams for each spaxel, as illustrated in Fig.~\ref{co_rotdiagram} for three positions. The CO lines in B1 are optically thin for $J\ge$9, as indicated by the $^{13}$CO lines detected with HIFI \citep{lefloch12}, thus the slope of the diagram gives directly 1/$T_{\rm ex}$.
We find a similar excitation temperature $T_{\rm{ex}}\simeq210\pm10$~K at all positions, indicating that the excitation conditions are fairly uniform, although at offset (9\arcsec,2\arcsec) the excitation temperature is slightly higher, 241 K. In fact, a secondary peak of H$_2\,v=$1-0 S(1) is observed at this position (see Fig.~\ref{comparison}) where \citet{nisini10}, from the H$_2$ lines, derived a temperature about 100 K higher than at the CO peak. The total column density of CO averaged over 13\farcs1, assuming LTE, ranges from 2$\times$10$^{15}$~\cmdue\ at (9\arcsec,2\arcsec) to 6$\times$10$^{15}$~\cmdue\ at ($-$5\arcsec,7\arcsec). 

\begin{figure}
\centering
\includegraphics[angle=-90,width=9cm]{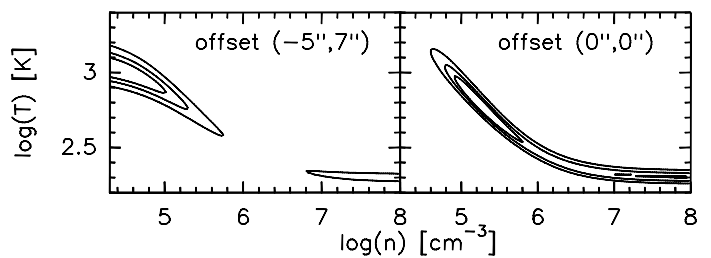}
\caption{$\chi^2$ distribution at the CO peak (left panel) considering only PACS lines and at the FOV centre (right panel) considering PACS and HIFI lines. Contours indicate 68, 90, and 99~\% confidence levels. }
\label{chisquare}
\end{figure}

Constraints on the kinetic temperature and density of the CO gas were obtained using the radiative transfer code RADEX \citep{vandertak07} in a plane-parallel geometry. We used the collisional rate coefficients with H$_2$ of \citet{yang10} and built a grid of models with density between 10$^4$~\cmtre\, and 10$^8$~\cmtre\, and temperature between 10~K and 10$^4$~K. In Fig.~\ref{chisquare} the region of minimum $\chi^2$ is shown as a function of density and temperature for the CO peak at ($-$5\arcsec,7\arcsec). The best fit is obtained for $n=$2$\times$10$^4$~\cmtre\, and $T_{\rm{kin}}$=1080~K, but other fits at lower temperatures combined with higher densities are also possible (see Fig. \ref{co_fit_3-2}), including the LTE solution.

Additional constraints on physical conditions are provided by the CO lines observed with HIFI towards the (0\arcsec,0\arcsec) position that, as shown by the rotational diagram, has the same excitation condition of offset ($-$5\arcsec,7\arcsec). Indeed, line profiles in CO\,(16--15) and CO\,(13--12) reveal a specific spectral shape for the excited CO component dominating the PACS range \citep{lefloch12}. Its flux contribution to CO\,(10--9) can then be extracted and scaled to 13\farcs1 by assuming the same spatial distribution as for CO\,(16--15).
The simultaneous fitting of the HIFI and PACS fluxes clearly excludes low-density / high-temperature solutions (Fig. \ref{chisquare} right panel). The best fit now has $T_{\rm{kin}}$=580~K and $n$=2$\times$10$^5$~\cmtre\, but still with a wide range of acceptable parameters, 200$<T_{\rm{kin}}<$1100~K for $n \geq 6\times 10^4$\cmtre\, (Fig. \ref{co_fit_2-2}). 

Tighter limits come from the comparison of the CO column density at (-5",7") inferred from RADEX fits with the H$_2$ column density at the same position and temperature derived from Spitzer data \citep{nisini10}, averaged over the same beam. We find that the ratio $N$(CO)/$N$(H$_2$) is (0.3--2)$\times10^{-4}$ for the best CO fits at $T_{\rm kin}$ of 200 and 580~K, respectively. In contrast, the CO relative abundance increases to unphysically high values for $T_{\rm kin}>$ 800~K. This clearly favours dense and moderately warm gas  (200$< T_{\rm kin}<$800~K, and $n\ge10^5$~\cmtre). 
From these best-fit models we calculate a total cooling of $L$(CO) = 0.004 \lsun\ at the CO peak and an intrinsic CO column density $N$(CO)= 2$\times$10$^{16}$ \cmdue, assuming the size of 7\arcsec\, derived from PACS maps. 

\subsection{[\ion{O}{i}] and OH lines: Evidence for a dissociative jet-shock}

\begin{figure}
\centering
\includegraphics[angle=90,width=8cm]{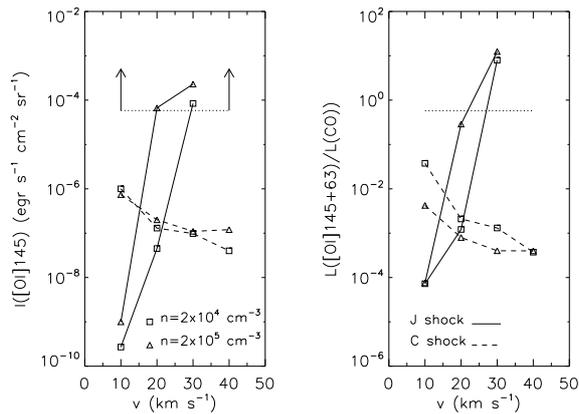}
\caption{Comparison between observed value (horizontal dotted line) and theoretical values for the \citet{flower10} shock models (connected symbols) of the  [\ion{O}{i}]145\um\, surface brightness
(left panel) and  the cooling ratio $L$([\ion{O}{i}]145+63)/$L$(CO) (right panel), as a function of shock speed. Solid lines refer to J-shocks and dashed lines to C-shocks; squares refer to pre-shock density $n$=2$\times$10$^4$ \cmtre\, and triangles to $n$=2$\times$10$^5$ \cmtre. }
\label{shock}
\end{figure}

In the left panel of Fig.~\ref{shock}, the surface brightness of the [\ion{O}{i}]145\um\, line (obtained using an upper limit to the size of 7\arcsec) is compared with shock-model predictions from \citet{flower10}. The [\ion{O}{i}]145/63 line ratio indicates that both lines are optically thin, but we used the [\ion{O}{i}]145\um\, line for this comparison, because the 63\um\ line may be affected by foreground absorption, as usually found by ISO \citep{liseau06}. Evidently only dissociative J-type shocks, with velocity $V_{\rm s} > 30$ \kms\, for $n$=2$\times$10$^4$ \cmtre\, and $V_{\rm s} > 20$ \kms\, for $n$=2$\times$10$^5$ \cmtre, can match the observed [\ion{O}{i}]145\um\, brightness. Non-dissociative shocks (of both J- and C-type) underpredict this line by at least two orders of magnitude. 
This conclusion is also supported by the detection of [\ion{Fe}{ii}] at 26 \um\, and [\ion{S}{i}] at 25 \um\, \citep{neufeld09} at the [\ion{O}{i}] peak; the iron and sulfur fine structure lines are excited only in dissociative shocks. In particular, the [\ion{O}{i}]/ [\ion{S}{i}] observed intensity ratio is consistent with both lines originating from a J-shock with the same conditions as inferred above \citep{hollenbach89}. The same models underpredict the [\ion{O}{i}]/[\ion{Fe}{ii}] but this is likely because the iron is partially depleted on grains.
Shock models reproducing the [\ion{O}{i}] brightness predict a ratio of CO to [\ion{O}{i}] total luminosity within a factor of a few of the observed one (see Fig. \ref{shock} right panel), therefore the dissociative J-shock traced by [\ion{O}{i}] may also contribute to the high-\jup\ CO emission.

The sum of the three OH  doublets at 79~\um, 84~\um, and 119~\um\ (see Table ~\ref{tab:OH}) amounts to three times the [\ion{O}{i}]145\um\ line, consistent with both species tracing the same dissociative J-shock \citep{hollenbach89}. Moreover, the ratio of OH to CO luminosity is an order of magnitude higher than predicted for C-shocks \citep{kaufman96}. Indeed, in C-shock all oxygen not locked in CO is converted in water, while in dissociative J-shocks the presence of FUV photons produced at the shock front delays the full conversion of free oxygen into water and maintains
a significant abundance of OH.
Interestingly, both the relative intensities of the OH doublets in L1157-B1 and the ratios of each OH doublet to [\ion{O}{i}]145\um\, and to $L_{\rm bol}=3 L_\odot$, are very similar to those observed much closer to low-mass protostars \citep{wampfler10,herczeg11}. This suggests that, OH may not be excited by infrared pumping, but predominantly by dissociative J-shocks in outflows far away as well as close to the source.

In dissociative J-type shocks, the [\ion{O}{i}]63\um\, luminosity is proportional to the mass flux into the shock, which is close to the wind mass-flux in a slow moving wind-shock \citep{hollenbach85}. Considering the total flux of the [\ion{O}{i}]63\um\, line in the PACS FOV ($10^{-15}$ W/m$^{-2}$), we derive  $L$([\ion{O}{i}]63\um) = 0.002 $L_\odot$ and $\dot M_{\rm w}$([\ion{O}{i}])= 2$\times$10$^{-7}$ \msun yr$^{-1}$ in the blue lobe. We may compare this value to independent estimate derived from the momentum and age of the blue outflow lobe traced by CO\,(1--0) by  \citet{bachiller01}. Rescaling the value these authors derived for a distance of 440 pc to our adopted distance of 250 pc, we infer $\dot M_{\rm w}$(CO) = 6$\times10^{-7}$ \msun yr$^{-1}$ under the assumption of outflow-wind momentum conservation and wind velocity of 100 \kms. Given the large uncertainties in the outflow age \citep{downes07} and possible self-absorption in the [\ion{O}{i}]63\um\, line, the two estimates are consistent and agree reasonably well with a jet-driven shock accelerating the outflow as the main excitation mechanism of [\ion{O}{i}] in B1.


\section{Conclusions}
In the Herschel-PACS spectrum of the B1 shocked region of the L1157 outflow, emission lines from CO, H$_2$O, OH and [\ion{O}{i}] have been detected. The analysis of the high-\jup CO lines with a radiative transfer code has allowed us to constrain the physical conditions of the warm (200$<T_{\rm{kin}}<800$~K) and dense ($n\geq10^5$~\cmtre) gas. The spatial resolution of PACS, unprecedented at FIR wavelengths, has shown that CO, OH and [\ion{O}{i}] all surprisingly peak at the rear of the bow shock, where others tracers of J-type shocks such as ro-vibrational H$_2$ and [\ion{Fe}{ii}] were also detected. The intensity of the bright [\ion{O}{i}] lines clearly indicates for the first time that a dissociative ``reverse'' wind-shock is the dominant exciting mechanism of these lines. The $L$(CO) and $L$(OH) are also consistent with a J-type shock.
Our analysis reveals that the C-shock previously found in B1 appears to give only a minor contribution to the emission of [OI], OH, and high-\jup CO lines detected in the PACS spectra.

\begin{acknowledgements}
PACS has been developed by a consortium of institutes led by MPE (Germany) and including UVIE (Austria); KU Leuven, CSL, IMEC (Belgium); CEA, LAM (France); MPIA (Germany); INAF-IFSI/OAA/OAP/OAT, LENS, SISSA (Italy); IAC (Spain). This development has been supported by the funding agencies BMVIT (Austria), ESA-PRODEX (Belgium), CEA/CNES (France), DLR (Germany), ASI/INAF (Italy), and CICYT/MCYT (Spain)
\end{acknowledgements}


\Online

\begin{appendix}

\section{}
\begin{figure}
\centering
\includegraphics[angle=-90,width=9cm]{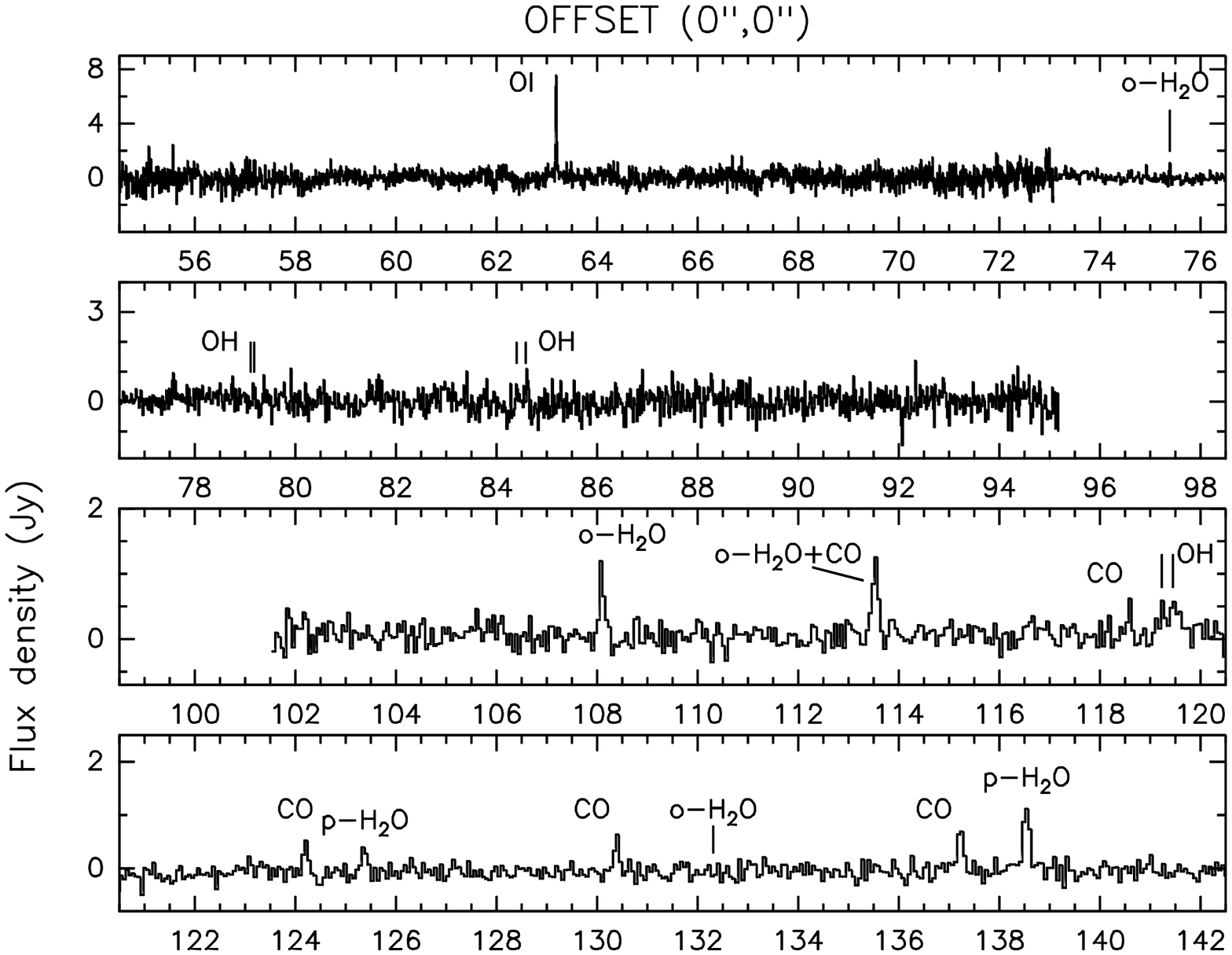}
\includegraphics[angle=-90,width=9cm]{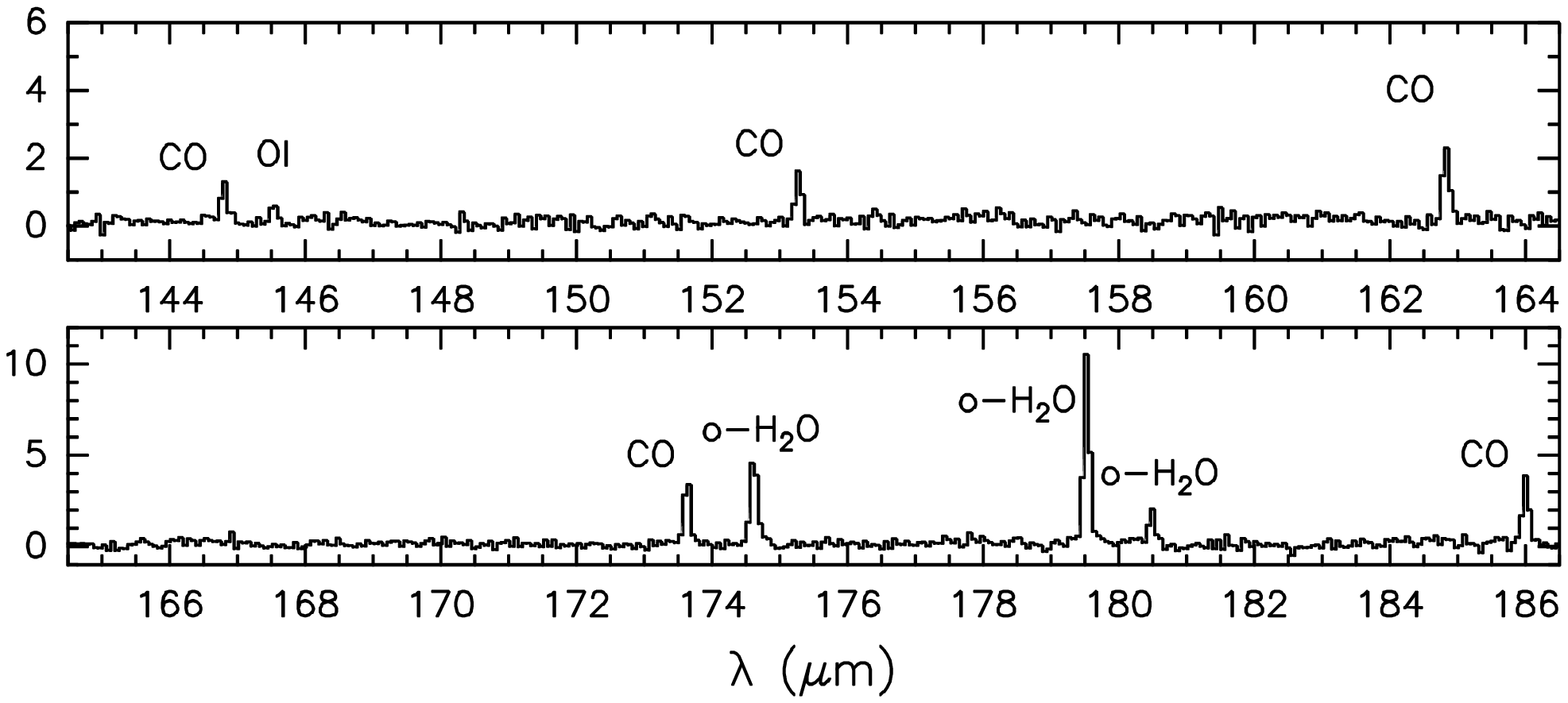}
\caption{Spectrum of the spatial pixel at the centre of the FOV, RA(J2000) = 20$^{\rm h}$39$^{\rm m}$10\fs2, Dec(J2000) = +68\degr01\arcmin10\farcs5.}
\label{spec2-2}
\end{figure}

\begin{figure}
\centering
\includegraphics[angle=-90,width=9cm]{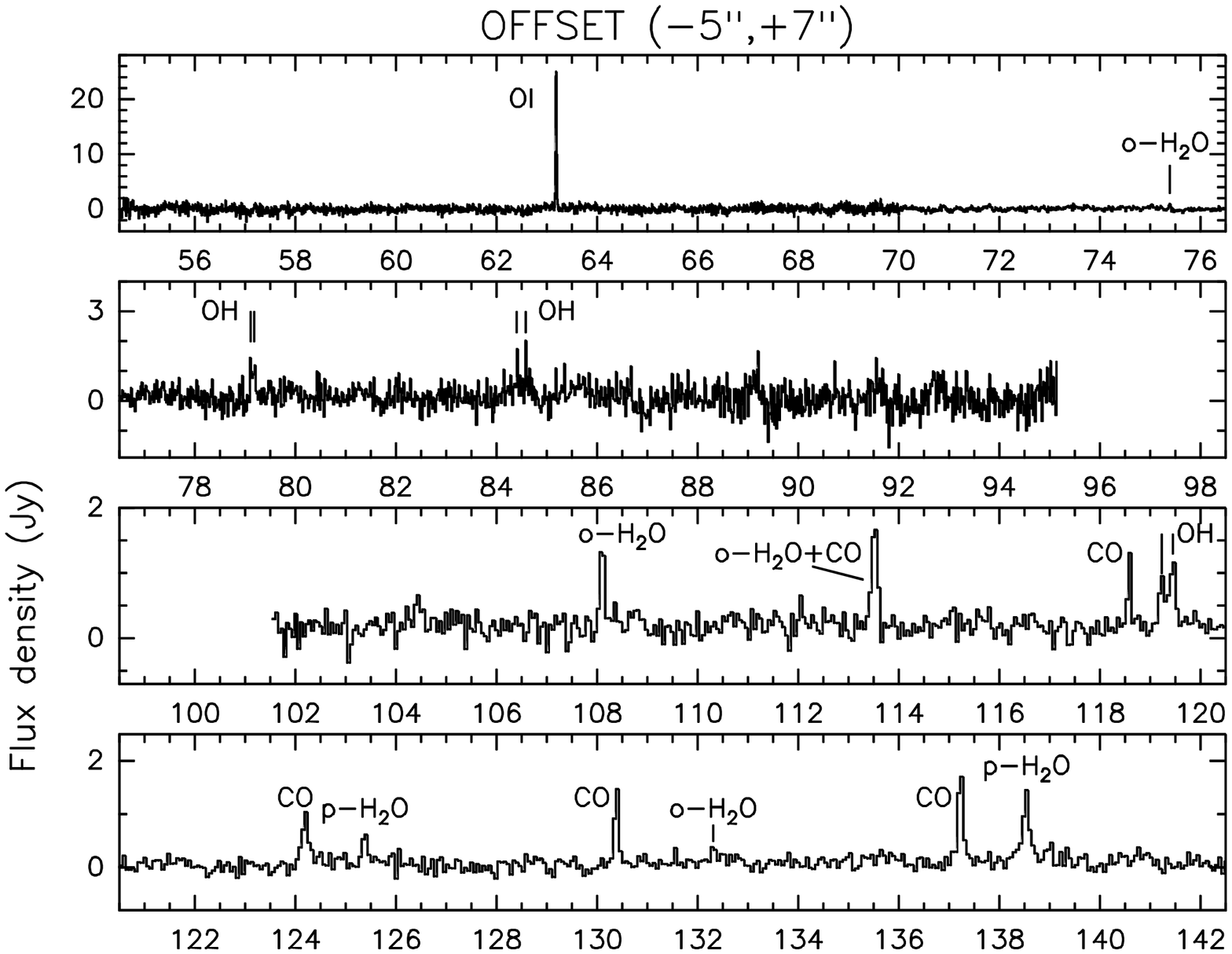}
\includegraphics[angle=-90,width=9cm]{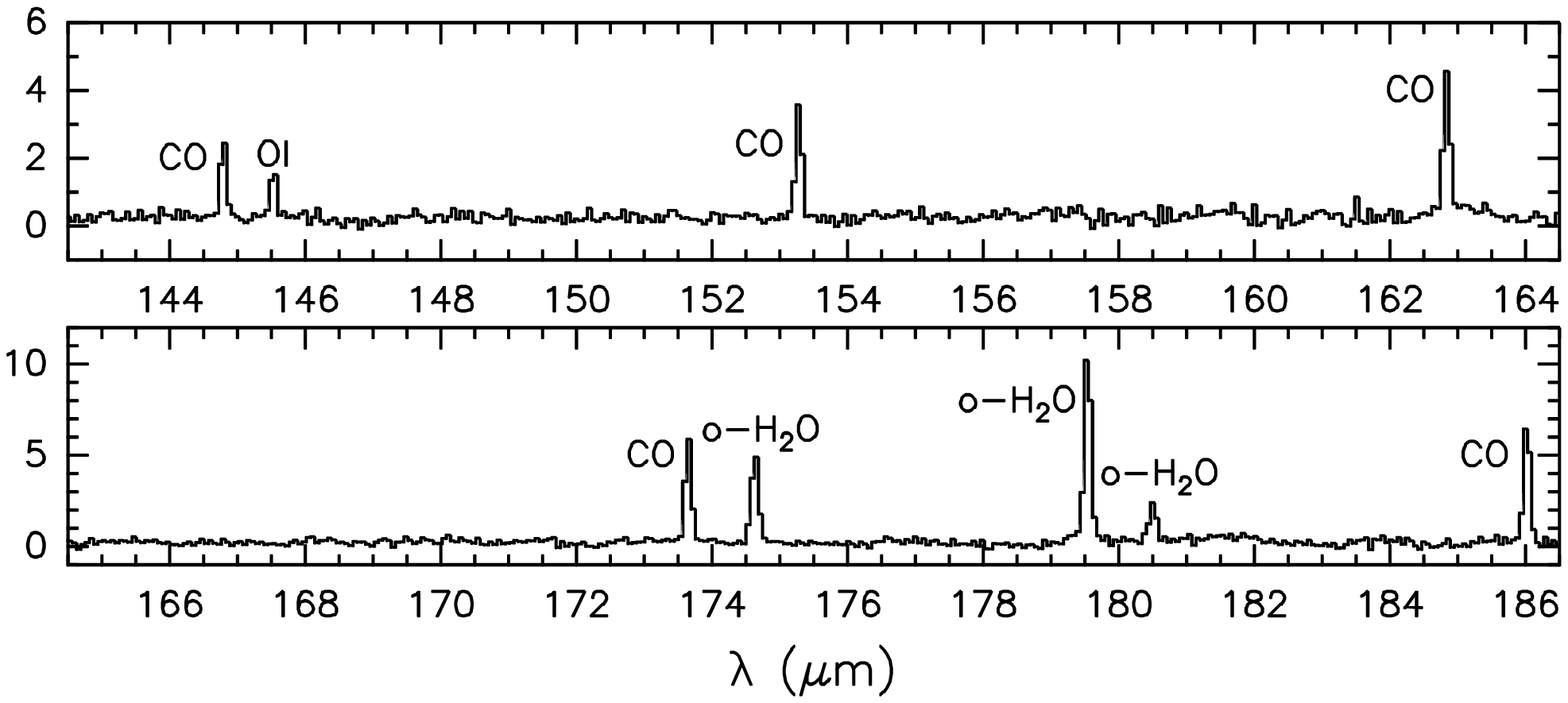}
\caption{Spectrum of the spatial pixel centred at RA(J2000) = 20$^{\rm h}$39$^{\rm m}$9\fs3, Dec(J2000) = +68\degr01\arcmin17\farcs8, corresponding to an offset ($-$5\arcsec,7\arcsec).}
\label{spec3-2}
\end{figure}

 In Fig. \ref{spec2-2} and \ref{spec3-2} we show the PACS spectra at the centre of the FOV (the nominal position of B1) RA(J2000) = 20$^{\rm h}$39$^{\rm m}$10\fs2, Dec(J2000) = +68\degr01\arcmin10\farcs5 and towards the PACS emission peak, at position RA(J2000) = 20$^{\rm h}$39$^{\rm m}$9\fs3, Dec(J2000) = +68\degr01\arcmin17\farcs8, located at offset ($-$5\arcsec; 7\arcsec) with respect to the centre of the FOV. 

In Table \ref{co_flux} we list the fluxes at offset (0\arcsec,0\arcsec) and ($-$5\arcsec,7\arcsec) extracted from the line maps convolved at the diffraction size of the CO transition at the highest wavelength, i.e. 13\farcs1 (PSF at 186 \um). They were obtained by multiplying the line surface brightness by an area of 1.133 $\times$ 13\farcs1$^2$, and were used for the rotational diagram and the RADEX modelling, assuming the same source area.  

\begin{table}
\caption{List of CO line fluxes at offset (0\arcsec,0\arcsec) and ($-$5\arcsec,7\arcsec),
derived from the line maps convolved at 13\farcs1 resolution.}
\label{flux_lvg}
\centering 
\begin{tabular}{|c|r|c|c|}
\hline
\hline
Line    &$\lambda_{rest}$ & Flux offset ($-$5\arcsec,7\arcsec) & Flux offset (0\arcsec,0\arcsec)\\
        &($\mu$m)   & (10$^{-21}$ W cm$^{-2}$) & (10$^{-21}$ W cm$^{-2}$)\\
\hline
CO (22\--21)  &118.58  & 3.2$\pm$0.5 & --\\
CO (21\--20)  &124.20  & 4.2$\pm$0.9 & 2.9$\pm$0.7 \\
CO (20\--19)  &130.39  & 4.3$\pm$0.6 & 3.0$\pm$0.5 \\
CO (19\--18)  &137.23  & 5.4$\pm$0.7 & 3.9$\pm$0.8 \\
CO (18\--17)  &144.79  & 6.3$\pm$0.8 & 4.5$\pm$0.6 \\
CO (17\--16)  &153.28  & 9.0$\pm$1.0 & 5.3$\pm$0.7 \\
CO (16\--15)  &162.82  &12.2$\pm$1.4 & 7.6$\pm$1.0 \\
CO (15\--14)  &173.63  &13.4$\pm$1.6 & 10.0$\pm$1.2 \\
CO (14\--13)  &186.01  &15.9$\pm$1.7 & 10.3$\pm$1.1 \\
CO (13\--12)  &220.25  & -- & 10.0$\pm$1.6 \\
CO (10\--9)   &260.24  & -- & 6.1$\pm$1.0 \\
\hline
\end{tabular}
\label{co_flux}
\end{table}

In Table \ref{tab:OH} we list the fluxes of the OH lines measured at offset ($-$5\arcsec,7\arcsec). Since the OH lines were detected only in one spaxel, we applied the flux correction for a point source that takes into account the size of the PSF.

\begin{table}
\centering
\caption{OH line fluxes measured at offset ($-$5\arcsec,7\arcsec) after PSF correction for a point
  source. The FWHM was set to the instrumental value.}
\begin{tabular}{|c|c|c|c|}
\hline
Line& $\lambda_{\rm rest}$ & FWHM & Flux\\
& ($\mu$m) & ($\mu$m) & (10$^{-17}$W/m$^2$) \\
\hline
OH $^2\Pi_{\frac{1}{2}}-^2\Pi_{\frac{3}{2}} J={\frac{1}{2}}^--{\frac{3}{2}}^+$ &79.11	&0.039& 3.5$\pm$0.9 \\
OH $^2\Pi_{\frac{1}{2}}-^2\Pi_{\frac{3}{2}} J={\frac{1}{2}}^+-{\frac{3}{2}}^-$ &79.18	&0.039& 3.1$\pm$0.9 \\
OH $^2\Pi_{\frac{3}{2}}-^2\Pi_{\frac{3}{2}} J={\frac{7}{2}}^+-{\frac{5}{2}}^-$ &84.42	&0.038& 3.1$\pm$0.9 \\
OH $^2\Pi_{\frac{3}{2}}-^2\Pi_{\frac{3}{2}} J={\frac{7}{2}}^--{\frac{3}{2}}^+$ &84.58	&0.038& 3.4$\pm$0.9 \\
OH $^2\Pi_{\frac{3}{2}}-^2\Pi_{\frac{3}{2}} J={\frac{5}{2}}^--{\frac{3}{2}}^+$ &119.23  &0.117& 3.0$\pm$0.4 \\
OH $^2\Pi_{\frac{3}{2}}-^2\Pi_{\frac{3}{2}} J={\frac{5}{2}}^+-{\frac{3}{2}}^-$ &119.44  &0.117& 4.2$\pm$0.4 \\        
OH $^2\Pi_{\frac{1}{2}}-^2\Pi_{\frac{1}{2}} J={\frac{3}{2}}^+-{\frac{1}{2}}^-$ &163.12  &0.126& $<$ 1.2 \\
OH $^2\Pi_{\frac{1}{2}}-^2\Pi_{\frac{1}{2}} J={\frac{3}{2}}^--{\frac{1}{2}}^+$ &163.40  &0.126& $<$ 1.2  \\
\hline 
\end{tabular}
\label{tab:OH}
\end{table}

\begin{figure}
\includegraphics[angle=0,width=9cm]{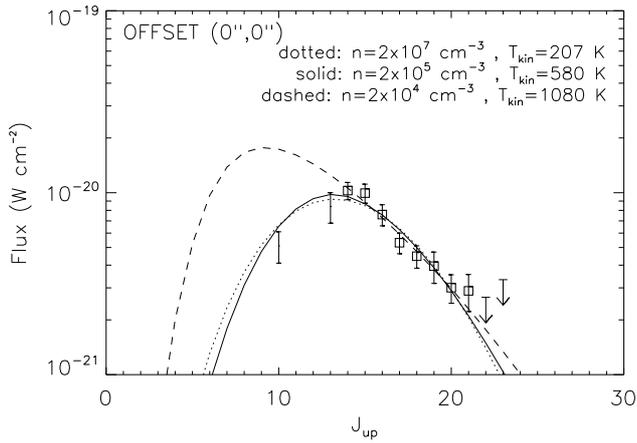}
\caption{Fit of the CO lines at offset (0\arcsec,0\arcsec), taking into account only PACS data (dashed curve), adding the HIFI lines (solid curve) and assuming the LTE temperature (dotted curve). Squares are the PACS fluxes and bars represent the range of fluxes estimated from HIFI lines. Error bars include the statistical error from the line fitting and the 10~\% calibration uncertainty.}
\label{co_fit_2-2}
\end{figure}

\begin{figure}
\includegraphics[angle=0,width=9cm]{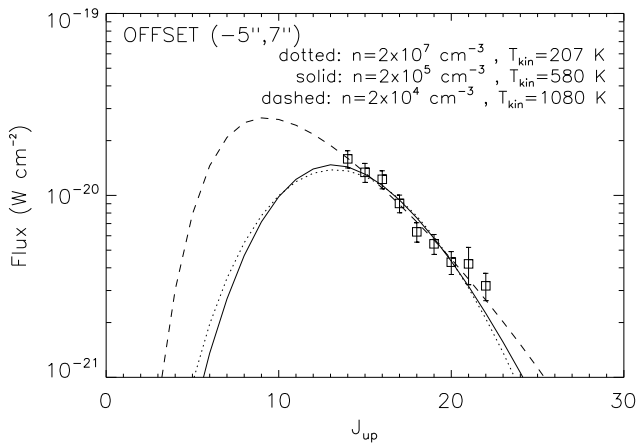}
\caption{Fit of the CO lines at offset ($-$5\arcsec,7\arcsec). Error bars include the statistical error from the line fitting and the 10~\% calibration uncertainty.}
\label{co_fit_3-2}
\end{figure}

In Fig. \ref{co_fit_2-2} we show the comparison between the CO line fluxes at the centre of the PACS FOV, offset (0\arcsec,0\arcsec), and three models: the best-fit model obtained considering only the PACS lines (\jup$\geqslant$14) with $n=$2$\times$10$^4$~\cmtre\, and $T_{\rm{kin}}=$1080~K; the best-fit model obtained considering also the HIFI CO\,(13--12) and (10--9) lines with $n=$2$\times$10$^5$~\cmtre\, and $T_{\rm{kin}}=$580~K and the LTE model with kinetic temperature equal to the excitation temperature derived from the rotational diagram, with $n=$2$\times$10$^7$~\cmtre\, and $T_{\rm{kin}}=$207~K. In Fig. \ref{co_fit_3-2} the same models are shown compared to the CO line fluxes at the peak of the CO emission, i.e. offset ($-$5\arcsec,7\arcsec).

\end{appendix}

\end{document}